\documentclass[conference]{IEEEtran}
\def\BibTeX{{\rm B\kern-.05em{\sc i\kern-.025em b}\kern-.08em
    T\kern-.1667em\lower.7ex\hbox{E}\kern-.125emX}}
\usepackage{amsmath}
\usepackage{amsthm}
\usepackage{amsfonts}
\usepackage{amssymb}
\usepackage{graphicx}
\usepackage{cite}
\usepackage{subfigure}
\usepackage{balance}
\usepackage{algorithm}
\usepackage{algorithmic}
\usepackage{enumerate}
\usepackage{color}
\usepackage{array}
\usepackage{indentfirst}
\usepackage{multirow}
\usepackage{booktabs}
\usepackage{color}
\usepackage{slashbox}
\usepackage{diagbox}
\usepackage{threeparttable}
\usepackage[svgnames,table]{xcolor}
\usepackage{bm}
\usepackage{bbm}
\usepackage{subfigure}
\usepackage{epstopdf}

\makeatletter

\newcommand{\Rmnum}[1]{\expandafter\@slowromancap\romannumeral #1@}
\newcommand{\tb}{\mathbf}
\makeatother

\begin{document}

\title{Deep CLSTM for Predictive Beamforming in Integrated Sensing and Communication-enabled Vehicular Networks }

\author{Chang Liu, Xuemeng Liu, Shuangyang Li, Weijie Yuan, Derrick Wing Kwan Ng}

\maketitle

{\bf\textit{Abstract---}Predictive beamforming design is an essential task in realizing high-mobility integrated sensing and communication (ISAC), which highly depends on the accuracy of the channel prediction (CP), i.e., predicting the angular parameters of users.
However, the performance of CP highly depends on the estimated historical channel stated information (CSI) with estimation errors, resulting in the performance degradation for most traditional CP methods.
To further improve the prediction accuracy, in this paper, we focus on the ISAC in vehicle networks and propose a convolutional long-short term (CLSTM) recurrent neural network (CLRNet) to predict the angle of vehicles for the design of predictive beamforming.
In the developed CLRNet, both the convolutional neural network (CNN) module and the LSTM module are adopted to exploit the spatial features and the temporal dependency from the estimated historical angles of vehicles to facilitate the angle prediction.
Finally, numerical results demonstrate that the developed CLRNet-based method is robust to the estimation error and can significantly outperform the state-of-the-art benchmarks, achieving an excellent sum-rate performance for ISAC systems.
\\[-1.5mm]

\textit{Keywords---}Integrated sensing and communication, predictive beamforming, deep learning, convolutional long-short term neural network, vehicular networks.}

\section{Introduction}

Future wireless networks are expected to provide not only heterogeneous connectivity but also highly accurate sensing capability \cite{liu2022integrated}. Conventionally, the communication and radar sensing are treated as two separated applications with different operating frequency bands \cite{liu2022survey}. However, with the rapid growth of the wireless technology, the frequency spectrum is experiencing severe congestion, which pushes the network provider to consider the harmonious coexistence of communication and radar sensing over the same frequency spectrum. Typically, there are two approaches to achieve this coexistence. One is to design powerful interference management algorithms that are able to suppress the interference generated from other functionalities, while the other one is to consider the integrated sensing and communication (ISAC) technology by applying a carefully designed waveform that jointly considers both communication and radar sensing functionalities. In particular, we will consider the latter approach to achieve the communication and radar sensing coexistence, because it has been widely acknowledged that ISAC technology is capable of providing performance improvements for both the functionalities.

Many recent advances in ISAC worth to be discussed. The waveform design for downlink ISAC was discussed in \cite{liu2018toward}, where several design criteria are considered. Specifically, the globally optimal solutions on the beampattern are obtained, which are further used to design low-complexity algorithms for achieving a desired tradeoff between the communication and radar performances.
The inter-pulse modulation was considered in ISAC aiming to detect the target using the mainlobe of the multiple-input multiple-output (MIMO) radar waveform, while transmitting information symbols using the sidelobes \cite{hassanien2015dual}. However, this type of ISAC transmission imposes strong assumption that the communication terminals must be within the sidelobe region to guarantee the information reception \cite{liu2018mu}. It should be noted that many of the existing works on ISAC leveraging the conventional orthogonal frequency-division multiplexing (OFDM) waveforms.
The orthogonal time frequency space (OTFS) modulation is recently proposed 2-dimensional (2D) modulation scheme that is designed in the so-called ``delay-Doppler (DD) domain'' \cite{wei2021orthogonal}. It has been acknowledged in the literature that the OTFS system enjoys a better communication performance than the OFDM system, especially over the high-mobility channels \cite{li2021performance}. Furthermore, the concept of DD domain aligns well with the conventional radar theory \cite{li2021hybrid, li2021cross}, where the delay and Doppler are two important parameters determining the target range and velocity. Therefore, the application of OTFS in ISAC transmissions is worth to be studied. Preliminary attempts on designing ISAC waveforms based on the OTFS signal have also available in the literature. In \cite{gaudio2020effectiveness}, the effectiveness of using OTFS signal for ISAC has been examined, in which the OTFS signals are shown to be capable of achieving a promising communication performance without sacrificing the radar sensing quality. A novel ISAC-empowered OTFS network was devised in~\cite{yuan2021integrated}. A key motivation of this work is that it bypasses the channel estimation process in the downlink by relying on the channel state information acquired from the radar sensing. A spatially spread OTFS modulation-based ISAC transmission scheme was discussed in \cite{li2022novel}, where the concept of spatially spreading was introduced to fully exploit the angular domain feature, resulting simple and insightful system models for both communication and radar sensing functionalities.

Although the concept of ISAC is appealing in theory, it faces many difficulties hindering its practical applications. Among those difficulties, the accurate angular prediction is one of the most important issue for implementing ISAC in practice.
Consider a base station (BS) equipping a monostatic radar as an example, where the ISAC signal is transmitted towards the communication terminals according to the angular prediction of the current time slot, while applying the angle estimation based on the radar echoes received at the coming time slot.
Consequently, the exact angular parameters at the current time slot cannot be known before the actual transmission in this case, and we must rely on powerful angular prediction algorithms for achieving good ISAC performance.
In \cite{liu2020radar}, an extended Kalman filtering (EKF) approach was adopted for angular prediction, which can obtain a good prediction performance at the cost of a high computational complexity.
Furthermore, a Bayesian inference-based angular prediction algorithm was appeared in \cite{yuan2020bayesian}, where a state evolution model was derived according to the geometric relationship between the motion parameters from the previous time slot to the upcoming time slot. Based on the state evolution model, a factor graph assisted Bayesian inference algorithm was proposed for angular prediction, where the required computational complexity is reduced compared to the EKF approach.
However, the aforementioned algorithms generally suffer from a limited prediction accuracy due to various practical reasons, such as the sophisticated target moving trajectory and the inaccurate modelling.

To further improve the prediction accuracy for realizing the predictive beamforming, in this paper, we focus on the ISAC-assisted vehicle-to-infrastructure (V2I) system taking into account the vehicle mobility and exploit the deep learning (DL) technology \cite{liu2020deepresidual} \cite{xie2020unsupervised, liu2020deeptransfer} to develop a predictive beamforming framework, where a convolutional long-short term (CLSTM) recurrent neural network (CLRNet) is first designed for angle prediction based on the historical estimated angles and a predicted angle-based beamforming design is then adopted to perform the predictive beamforming.
In particular, both the convolutional neural network (CNN) and the LSTM modules are adopted in CLRNet to exploit the spatial features and the temporal dependency from the historical estimated angles for improving the angle prediction accuracy.
Furthermore, through the neural network offline training, the well-trained CLRNet is able to extract robust features for angle prediction, which facilitates the design of an angle predictor robust to the historical estimation errors.
Finally, simulation results verify the efficiency of the developed predictive scheme and demonstrate that the proposed method can achieve a comparable performance with the perfect beamforming scheme based on the perfect real-time angles.

\emph{Notations}:
Superscripts $T$ and $H$ denote the transpose and the conjugate operations, respectively.
$\mathbb{C}$ and $\mathbb{R}$ are the sets of complex numbers and real numbers, respectively.
$\arccos x$ represents the inverse cosine operation of $x$.
$\mathcal{U}(a,b)$ is used to represent the uniform distribution within $[a,b]$.
${\mathcal{CN}}( \mu,\sigma^2 )$ is the circularly symmetric complex Gaussian (CSCG) distribution with $\mu$ and $\sigma^2$ being the mean value and the variance, respectively.
Similarly, ${\mathcal{CN}}( \mu,\sigma^2 )$ denotes the real-valued Gaussian distribution.
In addition, $\|\bm{x}\|$ and $|x|$ denote the $\ell_2$ norm and the absolute value of a vector $\bm{x}$ and a scalar $a$, respectively.

\section{System Model}
\begin{figure}[t]
  \centering
  \includegraphics[width=\linewidth]{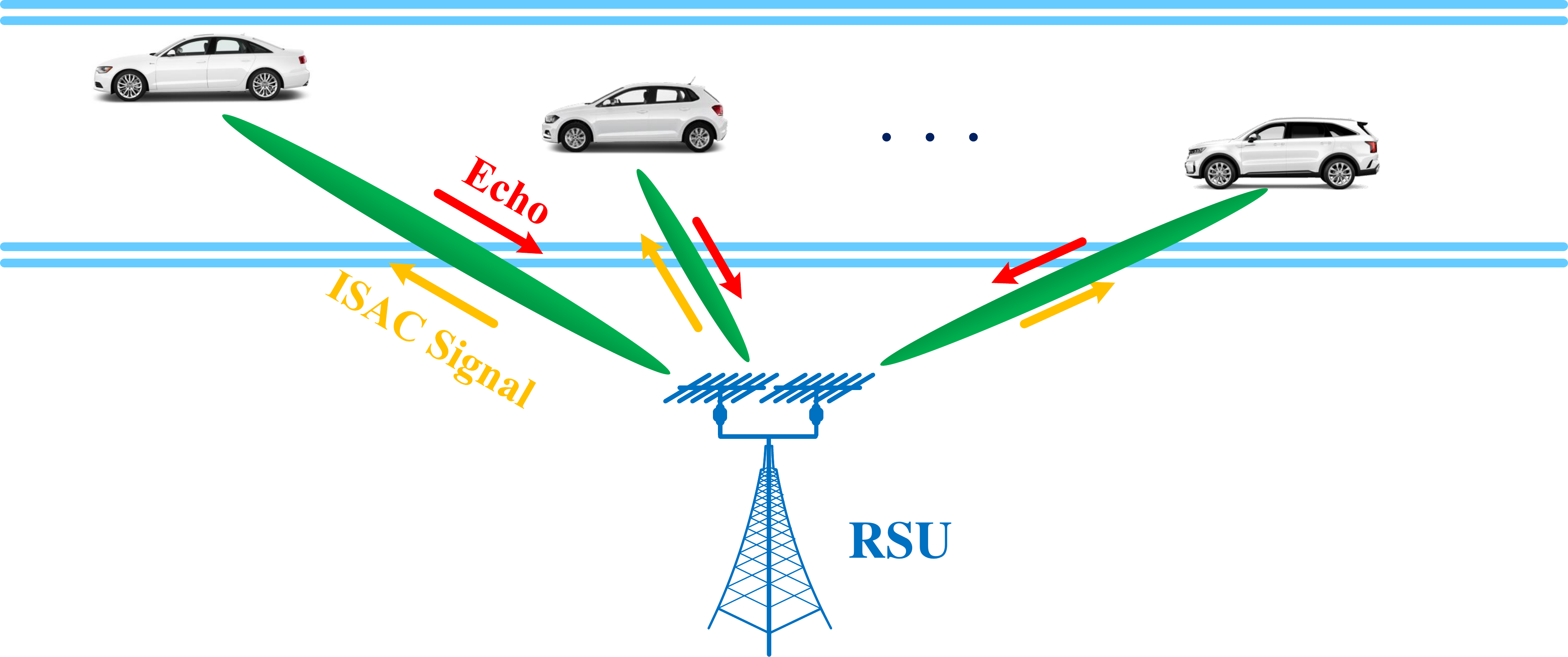}
  \caption{The considered ISAC-empowered V2I network.}\label{Fig:RSU_scenario}
\end{figure}

In this paper, an ISAC-empowered V2I system is considered, where we consider that a roadside unit (RSU) is deployed for serving $K$ single-antenna vehicles, as shown in Fig. \ref{Fig:RSU_scenario}.
The RSU is exactly an ISAC system, where a millimeter wave (mmWave)-based large scale uniform linear array (ULA) \cite{marzetta2016fundamentals} equipped with a $N_t$-transmit-antenna-array as well as a $N_r$-receive-antenna-array is deployed for signal transceiving.
In particular, a full-duplex radio technology \cite{barneto2021full} is adopted on ULA such that the RSU is able to simultaneously transmit the downlink signals and receive the sensing signals \cite{yuan2020bayesian}.

\subsection{Sensing Model}
Let $U_k$ denote the $k$-th, $k \in \mathcal{K} \triangleq\{1,2,\cdots,K\}$, vehicle and $s_{k,n}(t)$ denote the ISAC downlink signal for $U_k$ at time instant $t$ within the $n$-th, $n \in \{1,2,\cdots,N\}$, time slot.
In this case, the ISAC signal vector to be transmitted to all the $K$ vehicles can be expressed as
\begin{equation}\label{Wsn_t}
\tilde{\bm{s}}_n(t) = \bm{W}_n \bm{s}_{n}(t) \in \mathbb{C}^{N_t \times 1}.
\end{equation}
Here, $\bm{W}_n = [\bm{w}_{1,n},\bm{w}_{2,n},\cdots,\bm{w}_{K,n}]\in \mathbb{C}^{N_t\times K}$ represents the transmit beamforming matrix, where $\bm{w}_{k,n} \in \mathbb{C}^{N_t\times 1}$ denotes the beamforming vector with respect to $U_k$. Accordingly, the RSU receives the echo signals reflected from the vehicles, which is given by \cite{liu2020radar, liu2014maximum, yuan2020bayesian}
\begin{equation}
{\bm{r}}_n(t) =G \sum_{k=1}^K \beta_{k,n} e^{j2\pi \mu_{k,n}t} \bm{b}(\theta_{k,n}) \bm{a}^{\rm H}(\theta_{k,n})\tilde{\bm{s}}_{n}(t-\nu_{k,n}) + \bm{z}(t).
\end{equation}
Here, $\nu_{k,n}$ denotes the time-delay of $U_k$ and $\mu_{k,n}$ represents the Doppler frequency of $U_k$.
$G=\sqrt{N_t N_r}$ denotes the total antenna array gain.
In addition, $\theta_{k,n}$ represents the angle between $U_k$ and the RSU at time slot $n$ and $\beta_{k,n} = \frac{\varrho}{2d_{k,n}}$ denotes the reflection coefficient, with $\varrho$ and $d_{k,n}$ being the fading coefficient and the distance between $U_k$ and the RSU at time slot $n$, respectively.
Furthermore, $\tb{z}(t)\in \mathbb{C}^{N_r\times 1}$ represents a CSCG noise vector at the RSU.
It is worth noting that a line-of-sight (LoS) channel model \cite{Niu2015a} is usually adopted for mmWave communication systems, where the transmit steering vector and the receive steering vector \cite{zhao2018max, liu2016blind} at the RSU are formulated as \cite{zhao2018maximum, liu2014blind}
\begin{equation}\label{a}
  \bm{a}(\theta_{k,n})=\sqrt{\frac{1}{N_t}}[1,e^{-j\pi\cos\theta_{k,n}},\cdots,e^{-j\pi(N_t-1)\cos\theta_{k,n}}]^T
\end{equation}
and
\begin{equation}\label{b}
  \bm{b}(\theta_{k,n})=\sqrt{\frac{1}{N_r}}[1,e^{-j\pi\cos\theta_{k,n}},\cdots,e^{-j\pi(N_r-1)\cos\theta_{k,n}}]^T,
\end{equation}
respectively.
Due to the fact that the RSU is equipped with a massive MIMO system, the steering vectors associated with different angles can be regarded as asymptotically orthogonal \cite{marzetta2016fundamentals}, i.e.,
$\forall k \neq k^{'}$, $|\tb{b}^H(\theta_{k,n})\tb{b}(\theta_{k^{'},n})| \approx 0$ and $|\tb{a}^H(\theta_{k,n})\tb{a}(\theta_{k^{'},n})| \approx 0$.
Thus, different vehicles can be distinguished by RSU via different angle-of-arrivals (AoAs) such that the RSU can independently handle signals from each vehicle.
By exploiting some existing angle estimation algorithms, e.g., the AoA estimation algorithms \cite{Trees2004optimum, Gross2015smart} and the factor graph-based method \cite{yuan2020bayesian}, we can obtain the estimated angles from the received echo signals, as commonly adopted in e.g., \cite{yuan2020bayesian, liu2022learning}, which can be formulated as
\begin{equation}\label{theta_est}
  \theta_{k,n}^{\mathrm{E}} = \theta_{k,n} + \Delta \theta^{\mathrm{E}}_k
\end{equation}
where $\theta_{k,n}^{\mathrm{E}} \in \mathbb{R}$ represents the estimation of $\theta_{k,n}$ and $\Delta \theta^{\mathrm{E}}_k$ denotes the angle estimation error for $U_k$.
Generally, we assume $\Delta \theta^{\mathrm{E}}_k \sim \mathcal{N}(0,\sigma_E^2)$, $\forall k \in \mathcal{K}$, is a Gaussian random variable, where the variance $\sigma_E^2$ is set according to the normalized mean square error (NMSE) $\rho$, i.e., $\sigma_E^2 = \rho {E[\|\theta_{k,n}\|^2]}$.
In this case, stacking the estimated angles from all the $K$ vehicles into one vector, we can obtain the vector form of the estimated angles, i.e.,
\begin{equation}\label{theta_est_set}
  \Theta_{n}^{\mathrm{E}} = [\theta_{1,n}^{\mathrm{E}}, \theta_{2,n}^{\mathrm{E}}, \cdots, \theta_{K,n}^{\mathrm{E}}]^T \in \mathbb{R}^{K \times 1}.
\end{equation}


\subsection{Vehicle Mobility Model}
\begin{figure}[t]
  \centering
  \includegraphics[width=0.7\linewidth]{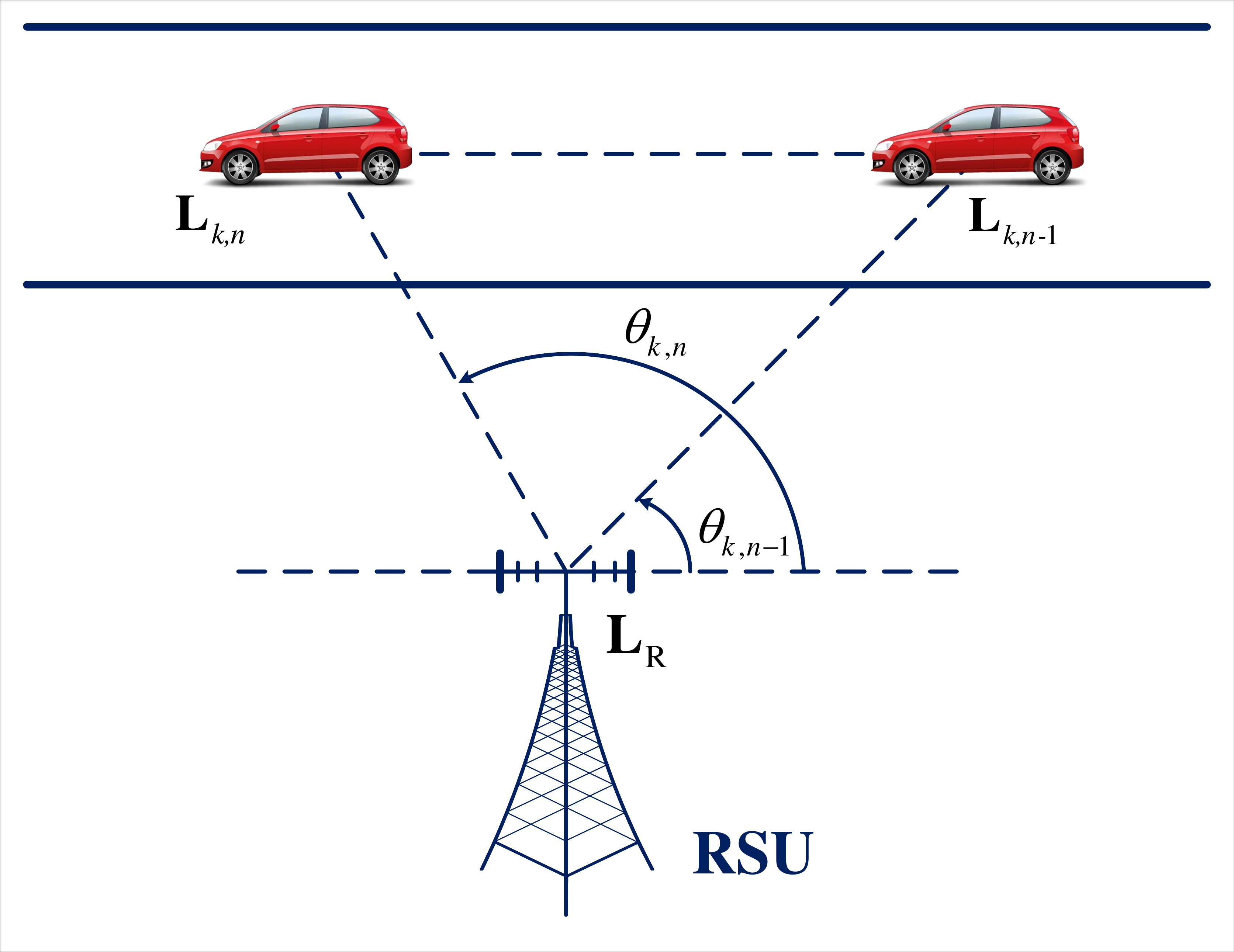}
  \caption{The vehicle mobility model in the considered ISAC-assisted V2I system.}\label{Fig:RSU_model}
\end{figure}

In this work, we focus on a high-mobility scenario is considered for ISAC systems, where the locations of vehicles change with time and the location of RSU is fixed.
As illustrated in Fig. \ref{Fig:RSU_model}, we adopt a two-dimensional (2D) coordinate to characterize the locations.
In particular, the RSU is located at $\bm{L}_\mathrm{R} = [x_\mathrm{R},y_\mathrm{R}]^T$, where $x_\mathrm{R}$ and $y_\mathrm{R}$ denote the coordinates of RSU on x-axis and y-axis, respectively. Similarly, the location of $U_k$ at time slot $n$ can be expressed as $\bm{L}_\mathrm{k,n} = [x_{k,n},y_{k,n}]$ with $x_{k,n}$ and $y_{k,n}$ being the coordinates of $U_k$ at time slot $n$ on x-axis and y-axis, respectively.
For ease of study, we assume that all vehicles keep in parallel to the road, as commonly adopted in e.g., \cite{wymeersch20175g, yuan2020bayesian, liu2022learning}. In this case, the angle of $U_k$ relative to the RSU at time slot $n$ can be formulated as
\begin{equation}\label{}
\theta_{k,n} = \arccos \frac{x_{k,n} - x_\mathrm{R}}{\|\bm{L}_\mathrm{k,n} - \bm{L}_\mathrm{R}\|},
\end{equation}
as illustrated in Fig. \ref{Fig:RSU_model}.
Thus, $\theta_{k,n}$ depends on the vehicle location and changes with $\bm{L}_\mathrm{k,n}$.
Without loss of generality, the kinematic equation of $U_k$ is given by \cite{liu2022learning, zeng2019axxessing}
\begin{equation}\label{}
  \bm{L}_{k,n} = \bm{L}_{k,n-1} + \bm{v}_{k,n-1}\Delta T + \bm{g}_{k,n-1},
\end{equation}
Here, $\bm{v}_{k,n-1} = [v_{k,n-1}^x, 0]$ denotes the average velocity of $U_k$ at time slot $n-1$ with $v_{k,n-1}^x$ being the velocity projections in the directions of $x$-axis.
In addition, $\Delta T$ denotes the time duration of each time slot.
Furthermore, $\bm{g}_{k,n-1} = [g_{k,n-1}^x, g_{k,n-1}^y] \sim \mathcal{N}(\bm{0},\sigma_g^2\bm{I})$ denotes a random vector to characterize the environment uncertainty of $U_k$ at time slot $n-1$, where $g_{k,n-1}^x$ and $g_{k,n-1}^y$ denote the uncertainty offsets at $x$-axis and $y$-axis, respectively, and $\sigma_g^2$ represents the offset variance.

\subsection{Communication Model}
In the downlink model, all the vehicles receive the transmitted signal from the RSU.
In particular, the received signal at $U_k$ can be formulated as
\begin{equation}\label{cmodel}
\vartheta_{k,n}(t) = \tilde{G} \sqrt{\alpha_{k,n}} e^{j2\pi \mu_{k,n}t}\tb{a}^{ H}(\theta_{k,n})\sum_{i=1}^K \bm{w}_{i,n}{s}_{i,n}(t) + \eta_{k,n}(t).
\end{equation}
Here, $\vartheta_{k,n}(t)$ is the received signal at the $t$-th time instant within the $n$-th time slot and $\tilde{G} = \sqrt{N_t}$ denotes the antenna gain.
In addition, $\alpha_{k,n} = \alpha_0(d_{k,n}/d_0)^{-\zeta}$ denotes the path loss with $\alpha_0$ and $\zeta$ being the reference path loss at $d_0$ and the path loss exponent, respectively.
Furthermore, $\eta_{k,n}(t) \sim \mathcal{CN}(0,\sigma_k^2)$ is adopted to represent the noise at $U_k$, where $\sigma_k^2$ denotes the received noise variance.
In this case, the received signal-to-interference-plus-noise ratio (SINR) at $U_k$ is formulated as
\begin{equation}\label{}
  \mathrm{SINR}_{k,n} = \frac{|\tilde{G}\sqrt{\alpha_{k,n}}\bm{a}^H(\theta_{k,n})\bm{w}_{k,n}|^2}
  {\sum_{j\in \mathcal{K}, j \neq k}|\tilde{G}\sqrt{\alpha_{k,n}}\bm{a}^H(\theta_{k,n})\bm{w}_{j,n}|^2 + \sigma_k^2}.
\end{equation}
As discussed in (\ref{a}), since a massive MIMO system is equipped with a large number of transmit antennas, the steering vectors associated with different angles can be regarded as asymptotically orthogonal, $\forall k \neq k^{'}$, we have $|\tb{a}^H(\theta_{k,n})\tb{a}(\theta_{k^{'},n})| \approx 0$. In this case, if we set $\bm{w}=\sqrt{p_{k,n}}\tb{a}(\theta_{k,n}^\mathrm{P})$, where $p_{k,n}$ denotes the signal power and $\theta_{k,n}^\mathrm{P}$ is the predictive value of $\theta_{k,n}$, the interference from other users is negligible.
Thus, the SINR is simplified to a signal-to-noise ratio (SNR), which is formulated as \begin{equation}\label{}
  \mathrm{SNR}_{k,n} (\theta_{k,n}^\mathrm{P}) = \frac{p_{k,n}|\tilde{G}\sqrt{\alpha_{k,n}}\bm{a}^H(\theta_{k,n})\tb{a}(\theta_{k,n}^\mathrm{P})|^2}
  {\sigma_k^2}.
\end{equation}
Correspondingly, the downlink sum-rate at time slot $n$ can be formulated as
\begin{equation}\label{R}
  R_n = \sum_{k=1}^{K} \log_2\left(1 + \mathrm{SNR}_{k,n} (\theta_{k,n}^\mathrm{P})\right).
\end{equation}
It is worth noting that $R_n$ depends on the value of $\theta_{k,n}^\mathrm{P}$, which can be obtained from the historical estimated angle values, i.e., $\theta_{k,n'}^\mathrm{P}$, $n'\leq n-1$, via the sensing function in ISAC system, as defined in (\ref{theta_est}).
However, the estimated angles are usually with estimation errors, which hinders the prediction accuracy of $\theta_{k,n}^\mathrm{P}$.
In the following, we will exploit the DL technology to develop a recurrent neural network-based approach robust to the estimation errors to facilitate the angle estimation for predictive beamforming.

\section{ The Proposed CLRNet-based Predictive Beamforming }
In this section, we will adopt a DL approach for angle prediction to realize the predictive beamforming in ISAC systems.
Without loss of generality, we assume that the estimated channel information within $\tau$ historical time slots can be obtained at the RSU for angle prediction.
In this case, if we specify that the $n-1$ time slot is the current time slot, then the task of angle prediction is to obtain
\begin{equation}\label{theta_pre_set}
  \Theta_{n}^{\mathrm{P}} = [\theta_{1,n}^{\mathrm{P}}, \theta_{2,n}^{\mathrm{P}}, \cdots, \theta_{K,n}^{\mathrm{P}}]^T \in \mathbb{R}^{K \times 1},
\end{equation}
based on the historical estimated channels, i.e., $\{\Theta_{n-1}^{\mathrm{E}}, \Theta_{n-2}^{\mathrm{E}},\cdots,\Theta_{n-\tau}^{\mathrm{E}}\}$, as defined in (\ref{theta_est_set}).
In the following, we will first introduce the developed neural network structure for predictive beamforming and then propose the DL-based predictive beamforming algorithm.

\subsection{Developed CLSTM-based Recurrent Neural Network (CLRNet)}
\begin{figure}[t]
  \centering
  \includegraphics[width=\linewidth]{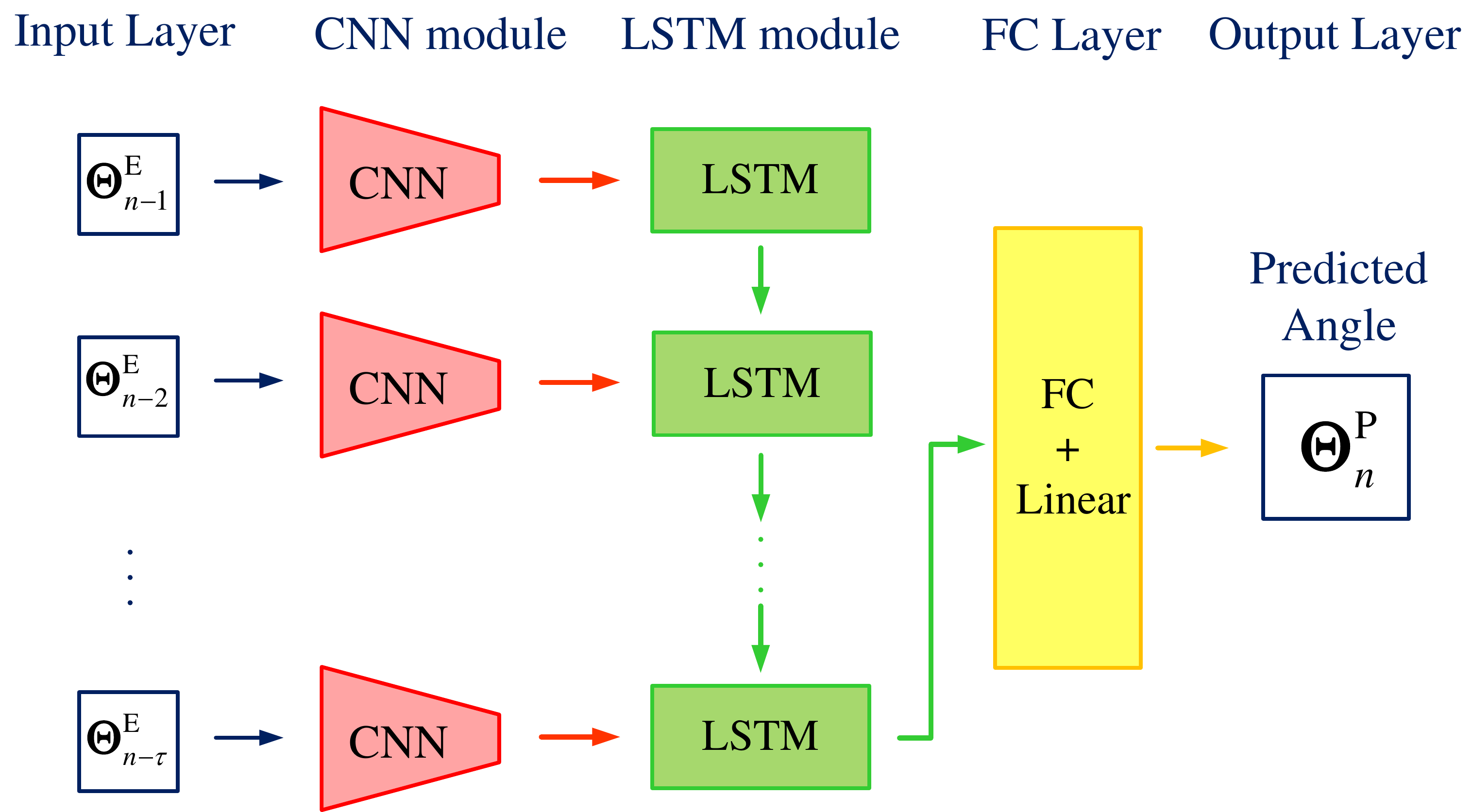}
  \caption{The developed CLRNet architecture for the predictive beamforming in the considered ISAC-assisted V2I network.}\label{Fig:CLSTM}
\end{figure}

Inspire by the powerful capability of the recurrent neural network in handling the sequential data, we develop a CLRNet to exploit both the spatial features and the temporal dependency from the historical estimated angles to predict the angle at the next time slot.
As illustrated in Fig. \ref{Fig:CLSTM}, the developed CLRNet consists of an input layer, one convolutional neural network (CNN) module, one LSTM module, one fully-connected (FC) layer, and one output layer.
The default hyperparameters of the CLRNet are summarized in Table \ref{Tab:Hyperparameters CLRNet}.
The details of the CLRNet will be introduce in the following.

\emph{a) Input Layer.}
Stacking the estimated angle vectors at different time slots into one matrix, we can obtain the set of the estimated angles from different vehicles at different time slots, which can be formulated as
\begin{equation}\label{Omega}
  \Omega_{n,\tau} \triangleq [\Theta_{n-1}^{\mathrm{E}}, \Theta_{n-2}^{\mathrm{E}}, \cdots, \Theta_{n-\tau}^{\mathrm{E}}].
\end{equation}
Since $\Omega_{n,\tau} \in \mathbb{R}^{K \times \tau}$ carries the temporal information of all the $K$ vehicles, we adopt it as the input of the CLRNet.

\emph{b) CNN Module.}
It is worth noting that the CLRNet is a recurrent neural network with $\tau$ time steps.
For each time step, we will first adopt a CNN module to extract features from the estimated angles at each time slot for the subsequent processing.
In particular, the CNN module consists of one convolutional layer and one flatten layer \cite{liu2019deep}.
The default filter sizes of different layers are set according to Table \ref{Tab:Hyperparameters CLRNet}, where ``conv. layer'' denotes the abbreviation of convolutional layer and a rectified linear unit (ReLU) is adopted after each convolution operation in the convolutional layer.

\emph{c) LSTM Module.}
For each time step, the LSTM module, i.e., an LSTM unit, is adopted after the CNN module to recurrently handle the extracted features at each time slot.
Specifically, for the previous $\tau-1$ time steps, the output of the LSTM unit in time step $\lambda$, $\lambda \in \{1,2,\cdots,\tau-1\}$, is sent to the LSTM unit for time step $\lambda + 1$.
Whereas, the output of the LSTM unit for the last time step $\tau$ is regarded as the LSTM module final output for it carries the temporal features extracted from estimated angles within the past $\tau$ time slots.

\emph{d) FC Layer.}
Given the extracted features from previous layers/modules, we adopt an FC layer with a linear activation function to make full use of these features for finally generating the desired output.

\emph{e) Output Layer.}
Based on the processing from a) to d), the CLRNet can finally outputs the predicted angle at time slot $n$, i.e., $\Theta_{n}^{\mathrm{P}}$, as defined in (\ref{theta_pre_set}).

According to the above discussions, the predicted angle generated by the CLRNet can be formulated as
\begin{equation}\label{CLRNet}
  \Theta_{n}^{\mathrm{P}} = f_\omega (\Omega_{n,\tau}),
\end{equation}
where $f_\omega (\cdot)$ denotes the mathematical expression of the developed CLRNet with $\omega$ being the neural network parameters.

Note that once the system parameters, e.g., $K$, $\tau$, change, the structure of the developed CLRNet can be easily refined by changing the sizes of the neural network accordingly.
Thus, the developed CLRNet is scalable and can be adopted for systems with different deployments.

\begin{table}[t]
\normalsize
\caption{Default hyperparameters of the developed CLRNet}\label{Tab:Hyperparameters CLRNet}
\centering
\small
\renewcommand{\arraystretch}{1.25}
\begin{tabular}{c c c}
  \hline
   \multicolumn{3}{l}{\textbf{Input Layer}: $\Omega_{n,\tau} = [\Theta_{n-1}^{\mathrm{E}}, \Theta_{n-2}^{\mathrm{E}}, \cdots, \Theta_{n-\tau}^{\mathrm{E}}] \in \mathbb{R}^{K \times \tau}$ }  \\
  \hline
   \textbf{Modules / Layers} & \textbf{Parameters} &  \hspace{0.3cm} \textbf{Values}   \\
   CNN module - Conv. layer & Filter size & \hspace{0.3cm}  $ 2 \times 2 $   \\
   LSTM module & Output size  & \hspace{0.3cm}  $ 8 \times 1 $   \\
   FC layer & Activation function & \hspace{0.3cm} Linear \\
  \hline
   \multicolumn{3}{l}{\textbf{Output Layer}: $\Theta_{n}^{\mathrm{P}} \in \mathbb{R}^{K \times 1}$ }   \\
  \hline
\end{tabular}
\end{table}

\subsection{Proposed CLRNet-based Predictive Beamforming Algorithm}
Based on the developed CLRNet, we can then propose a CLRNet-based predictive beamforming algorithm, which consists of the offline neural network training and the online predictive beamforming.
In the following, we will introduce the details of the proposed algorithm.

\emph{a) Offline Neural Network Training.}
Denote by
\begin{equation}\label{training_dataset}
  (\mathcal{X}, \mathcal{Y}) \triangleq \{ (\Omega_{n,\tau}^{(1)}, \Theta_n^{(1)}), (\Omega_{n,\tau}^{(2)}, \Theta_n^{(2)}), \cdots, (\Omega_{n,\tau}^{(N_t)}, \Theta_n^{(N_t)}) \}
\end{equation}
the training dataset. Here, $(\Omega_{n,\tau}^{(i)}, \Theta_n^{(i)})$ represents the $i$-th, $i\in \{1,2,\cdots,N_t\}$, offline training example of $(\mathcal{X}, \mathcal{Y})$, where $\Omega_{n,\tau}^{(i)}$ denotes the neural network input of the $i$-th training example, as defined in (\ref{Omega}) and $\Theta_n^{(i)} = [\theta_{1,n}^{(i)}, \theta_{2,n}^{(i)}, \cdots, \theta_{K,n}^{(i)}]^T \in \mathbb{R}^{K \times 1}$ denotes the label of the $i$-th training example, i.e., the truth values of angles.
According to the defined task in (\ref{theta_pre_set}), the cost function of the developed CLRNet can be formulated as a mean square error function \cite{Goodfellow2016deep, liu2021location, lxm2021deep, xie2019activity}, which can be formulated as
\begin{equation}\label{cost_function}
  J_{\mathrm{CLRNet}}(\omega) = \frac{1}{2N_t}\sum_{i=1}^{N_t} \left\| \Theta_n^{(i)} - f_\omega (\Omega_{n,\tau}^{(i)})\right\|^2.
\end{equation}
Given the training dataset in (\ref{training_dataset}), we can then minimize the cost function in (\ref{cost_function}) via the backpropagation algorithm (BPA) to obtain a well-trained CLRNet, i.e., $f_{\omega^*}(\cdot)$ with
\begin{equation}\label{well_trained_CLRNet}
  \omega^* = \min_{\omega}~J_{\mathrm{CLRNet}}(\omega)
\end{equation}
being the well-trained neural network parameters.

\emph{b) Online Predictive Beamforming.}
After offline neural network training, we can directly adopt the well-trained CLRNet to generate the predicted angle based on the historical estimated angles, which can be expressed as
\begin{equation}\label{online_test}
  \dot{\Theta}_n^{\mathrm{P}} = f_{\omega^*}(\dot{\Omega}_{n,\tau})
\end{equation}
where $\dot{\Theta}_n^{\mathrm{P}}$ denotes the predicted angle based on the well-trained CLRNet $f_{\omega^*}(\cdot)$ and the test example $\dot{\Omega}_{n,\tau}$.
Then, based on the predicted angle, we can design the predictive beamforming vectors for all the $K$ vehicles, i.e.,
\begin{equation}\label{}
  \bm{w}_{k,n} = \bm{a}(\dot{\Theta}_n^{\mathrm{P}}[k]), \forall k \in \mathcal{K}.
\end{equation}

\begin{table}[t]
\small
\centering
\begin{tabular}{l}
\toprule[1.8pt] \vspace{-0.3 cm}\\
\hspace{-0.1cm} \textbf{Algorithm 1} {CLRNet-based Predictive Beamforming Algorithm} \vspace{0.2 cm} \\
\toprule[1.8pt] \vspace{-0.3 cm}\\
\textbf{Initialization:} The training dataset $(\mathcal{X},\mathcal{Y})$ and $i_t = 0$ \\
\textbf{Offline Neural Network Training:} \\
1:\hspace{0.75cm}\textbf{Input:} Training dataset $(\mathcal{X},\mathcal{Y})$\\
2:\hspace{1.1cm}\textbf{while} $i_t \leq N_{\max} $ \textbf{do} \\
3:\hspace{1.6cm}Update $\omega$ via BPA to minimize $J_{\mathrm{CLRNet}}(\omega)$ in (\ref{cost_function}) \\
\hspace{1.8cm} $i_t = i_t + 1$  \\
4:\hspace{1.1cm}\textbf{end while} \\
5:\hspace{0.75cm}\textbf{Output}:  The well-trained CLRNet $f_{\omega^*}( \cdot ) $ in (\ref{well_trained_CLRNet})\\
\textbf{Online Predictive Beamforming:} \\
6:\hspace{0.75cm}\textbf{Input:} Test example $\dot{\Omega}_{n,\tau}$ in (\ref{online_test}) \\
7:\hspace{1.1cm}\textbf{do} Angle Prediction using $f_{\omega^*}( \cdot ) $ \\
8:\hspace{0.75cm}\textbf{Output:} Predicted angles $\dot{\Theta}_n^{\mathrm{P}} = f_{\omega^*}(\dot{\Omega}_{n,\tau})$ \\
9:\hspace{1.1cm}\textbf{do} Predictive Beamforming based on $\dot{\Theta}_n^{\mathrm{P}}$ \\
10:\hspace{0.6cm}\textbf{Output:} $\bm{w}_{k,n} = \bm{a}(\dot{\Theta}_n^{\mathrm{P}}[k]), \forall k \in \mathcal{K}$ \vspace{0.2cm}\\
\bottomrule[1.8pt]
\end{tabular}
\end{table}

\emph{c) Proposed Algorithm Steps.}
According to the above discussions, we then propose the CLRNet-based predictive beamforming algorithm and the associated algorithm steps are summarized in \textbf{Algorithm 1}, where $i_t$ and $N_{\max}$ are the iteration index and the maximum iteration number of the offline neural network training, respectively.

\section{Simulation Results}
In this section, we present simulation results in the ISAC-assisted V2I mmWave systems to verify the effectiveness of the developed predictive beamforming algorithm.
As introduced in Fig. \ref{Fig:RSU_scenario}, we consider a classical V2I system, where one RSU is equipped with $N_t = 32$ transmit antennas and $N_r = 32$ receive antennas for serving $K = 8$ single-antenna vehicles.
Moreover, we adopt a 2D coordinate to characterize the mobilities of the vehicles, as illustrated in Fig. \ref{Fig:RSU_model}. For breviety, the RSU is set at the location of $[0~\mathrm{m},0~\mathrm{m}]$ and the initial locations of all the vehicles are assumed to be random, i.e., $\forall U_k$, the associated initial location is set as $[x_{k}^{\mathrm{I}},y_{U_k}^{\mathrm{I}}] = [\bar{x}_{k},\bar{y}_{k}] + [\Delta x, \Delta y]$.
Here, $[x_{k}^{\mathrm{I}}, y_{k}^{\mathrm{I}}]$ represents the initial coordinate of $U_k$ and $[\bar{x}_{k},\bar{y}_{k}]$ denotes the mean value of the initial coordinate with a default setting of $[\bar{x}_{k},\bar{y}_{k}]=[25~\mathrm{m}, 10~\mathrm{m}]$.
In addition, $\Delta x, \Delta y$ are assumed to be Gaussian random variables with $\Delta x, \Delta y \in \mathcal{N}(0,1)$ to characterize the randomness of $[x_{k}^{\mathrm{I}},y_{U_k}^{\mathrm{I}}]$.
For the vehicle mobility model, we set $\Delta T = 0.02~\mathrm{s}$ and $\forall k,n$, the average velocity is randomly generated with $v_{k,n}^x \sim \mathcal{U}(8~\mathrm{m/s}, 8.25~\mathrm{m/s})$, i.e., the average velocity of each vehicle is set around $30~\mathrm{km/h}$.
Furthermore, a path loss model is adopted to characterize the large-scale fading, i.e., $\alpha_{k,n} = \alpha_0(d_{k,n}/d_0)^{-\zeta}$, as defined in (\ref{cmodel}), where we set $\alpha_0 = -65~\mathrm{dB}$, $d_0 = 1~\mathrm{m}$, and $\zeta = 3$ \cite{Niu2015a, liu2019maximum, liu2017optimal, hu2021robust, li2022how, ye2021delay}.
Also, $\forall k,n$, the noise variance is set as $\sigma_k^2 = -80~\mathrm{dBm}$ and the transmit power for $U_k$ is set as $p_{k,n} = P/K$, where $P$ denotes the total transmit power.
For the proposed predictive beamforming scheme, we assume $\tau = 6$, $N_t = 10,000$, and the associated neural network hyperparameters are set according to Table 1.

To evaluate the beamforming performance, two benchmarks are adopted as introduced in the following.
\begin{enumerate}[(a)]
  \item Perfect beamforming: In this scheme, the real-time actual angles of vehicles are adopted to realize the perfect beamforming design.
  \item Model-based method: This scheme is based on the state evolution model, i.e., $\sin(\theta_{k,n}^{\mathrm{P}}-\theta_{k,n-1}^{\mathrm{E}})d_{k,n-1} = v_{k,n-1}\Delta T \sin \theta_{k,n-1}^{\mathrm{E}}$, which can be easily derived from the system model in Fig. \ref{Fig:RSU_model}.
      Thus, the predicted angle can be expressed as \cite{liu2022learning}
      \begin{equation}\label{}
        \theta_{k,n}^{\mathrm{P}} = \arcsin\left(\frac{v_{k,n-1}\Delta T \sin\theta_{k,n-1}^{\mathrm{E}}}{d_{k,n-1}}\right) + \theta_{k,n-1}^{\mathrm{E}}, \forall k,n,
      \end{equation}
      where for ease of study, we only consider the historical angles with estimated errors and ideally assume that the other parameters, e.g., the velocities and the distances, are perfect known.
\end{enumerate}

Furthermore, all the points in the presented numerical results are generated by the mean results of $2,000$ Monte Carlo realizations.

\begin{figure}[t]
  \centering
  \includegraphics[width=3in,height=2.6in]{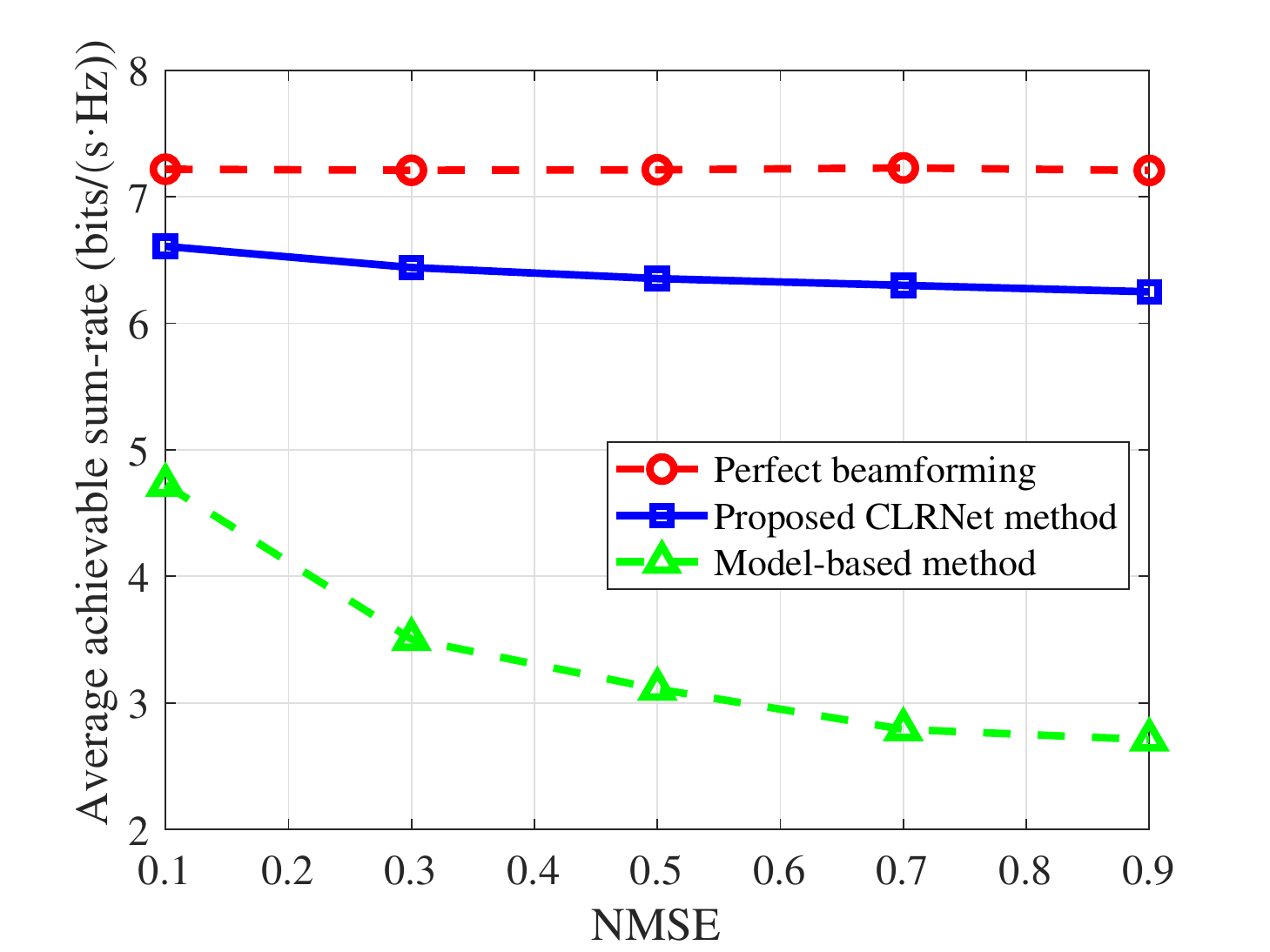}
  \caption{The average achievable sum-rate versus the NMSE under $P=20~\mathrm{dBm}$.}\label{Fig:rate_nmse}
\end{figure}

We first investigate the impact of the estimation errors of the historical channels on the beamforming performance.
As shown in Fig. \ref{Fig:rate_nmse}, we fix the total transmit power at $P=20~\mathrm{dBm}$ and evaluate the average achievable  sum-rate performance via varying the NMSE values of the historical angle estimation, as defined in (\ref{theta_est}).
It can be observed that with the increase of the NMSE, the performance of the model-based method drops seriously compered with the performance of the perfect beamforming scheme.
This is because in the model-based method, the estimated historical angles with a large NMSE are adopted for angle prediction, which leads to a low angle prediction accuracy for beamforming design and thus results in the system performance degradation.
In contrast, when the NMSE increases, our proposed CLRNet-based predictive beamforming method only has a slight performance drop and can achievable system performance compared with the perfect beamforming scheme.
This is expected since the developed method can not only exploit the temporal dependency from the historical information to improve the angle prediction accuracy, but also learn robust features from the inaccurate observations for angle prediction to grant a satisfactory system performance under different NMSEs.

\begin{figure}[t]
  \centering
  \includegraphics[width=3in,height=2.6in]{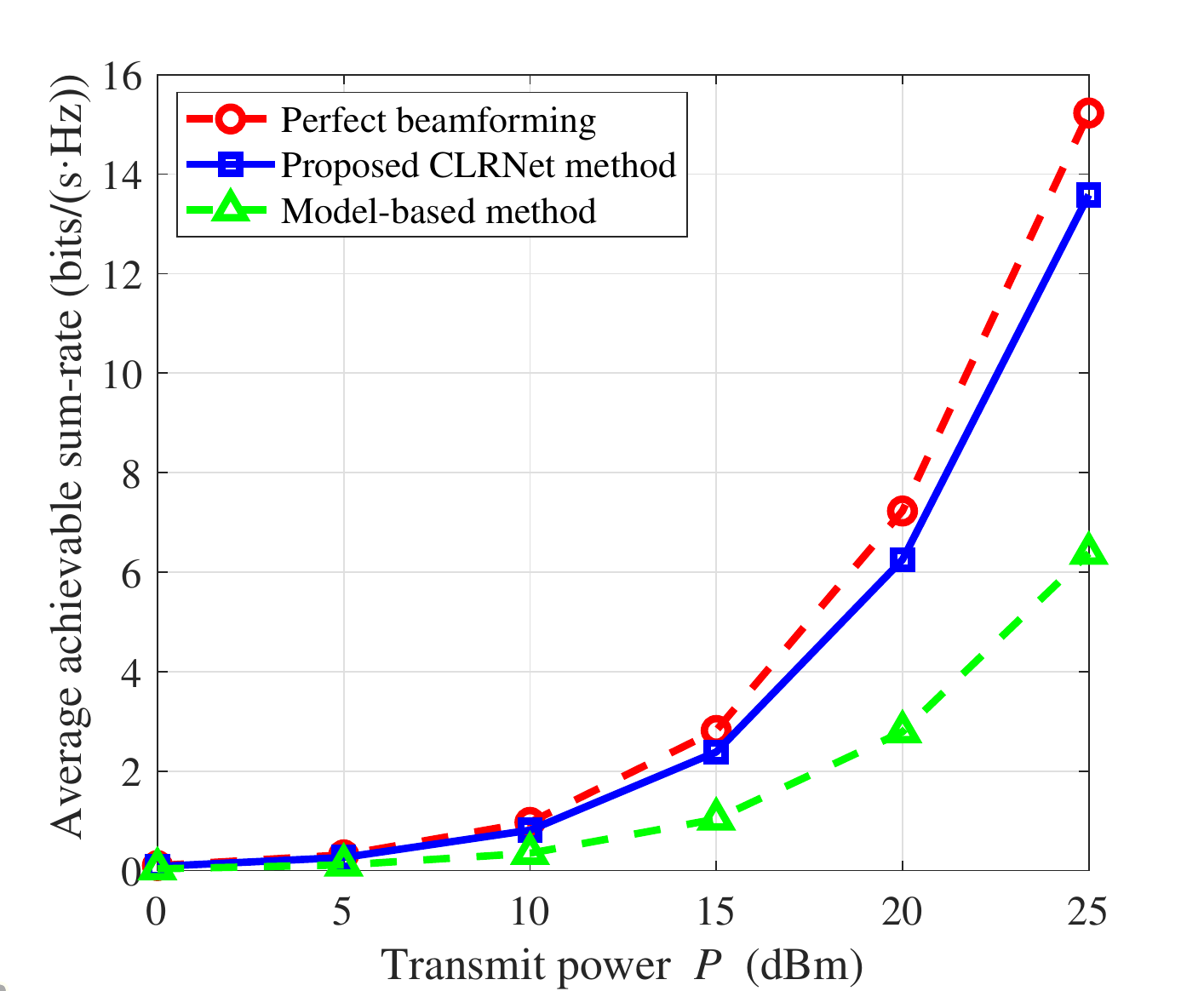}
  \caption{The average achievable sum-rate versus the total transmit power under the $\mathrm{NMSE}= 0.7$.}\label{Fig:rate_power}
\end{figure}

Correspondingly, we then present the curves of the average sum-rate versus the total transmit power in Fig. \ref{Fig:rate_power}, where the NMSE of the angle estimation is set as $\mathrm{NMSE} = 0.7$ and $P$ varies within $[0~\mathrm{dBm},25~\mathrm{dBm}]$.
We can find that the sum-rate of all the considered methods increase with the transmit power.
In particular, the performance of the proposed method almost increases with the same slope as the perfect beamforming scheme and the proposed method outperforms the model-based method significantly, e.g., achieving a $5~\mathrm{dB}$ performance gain at the sun-rate of $6~\mathrm{bits/s/Hz}$.
The reason is that the developed method can adopt the CLSTM module to extract spatial and temporal features from historical estimated angles to further improve the angle prediction accuracy for beamforming design.

\section{Conclusions}
This paper investigated the ISAC-assisted V2I systems and adopted a DL approach to predict angles of vehicles for beamforming design.
In the developed scheme, we proposed a predictive beamforming framework, where a CLRNet is first designed for angle prediction based on the historical estimated angles and a predicted angle-based beamforming design is then adopted to realize the ISAC functions.
In CLRNet, a CLSTM module is specifically adopted to exploit the spatial and temporal features to further improve the learning performance.
Meanwhile, by exploiting the powerful data-driven capability, the developed CLRNet can be trained as a robust predictor, which is hardly affected by the estimation errors and thus grants an excellent system performance.
Finally, simulation results demonstrated that the performance of the proposed predictive method can approach that of the perfect beamforming scheme requiring the availability of the real-time actual angles.

\bibliographystyle{ieeetr}

\setlength{\baselineskip}{10pt}

\bibliography{ReferenceSCI2}

\begin{thebibliography}{10}

\bibitem{liu2022integrated}
F.~Liu, Y.~Cui, C.~Masouros, J.~Xu, T.~X. Han, Y.~C. Eldar, and S.~Buzzi,
  ``Integrated sensing and communications: Towards dual-functional wireless
  networks for 6g and beyond,'' {\em IEEE J.\ Sel.\ Areas\ Commun.}, vol.~40,
  no.~6, pp.~1728--1767, Jun. 2022.

\bibitem{liu2022survey}
A.~Liu {\em et~al.}, ``A survey on fundamental limits of integrated sensing and
  communication,'' {\em IEEE Commun. Surveys Tuts.}, vol.~24, no.~2,
  pp.~994--1034, 2nd Quat., 2022.

\bibitem{liu2018toward}
F.~Liu, L.~Zhou, C.~Masouros, A.~Li, W.~Luo, and A.~Petropulu, ``Toward
  dual-functional radar-communication systems: Optimal waveform design,'' {\em
  IEEE Transactions on Signal Processing}, vol.~66, no.~16, pp.~4264--4279,
  Aug. 2018.

\bibitem{hassanien2015dual}
A.~Hassanien, M.~G. Amin, Y.~D. Zhang, and F.~Ahmad, ``Dual-function
  radar-communications: Information embedding using sidelobe control and
  waveform diversity,'' {\em IEEE Trans.\ Signal Proc.}, vol.~64, no.~8,
  pp.~2168--2181, Apr. 2016.

\bibitem{liu2018mu}
F.~Liu, C.~Masouros, A.~Li, H.~Sun, and L.~Hanzo, ``{MU-MIMO communications
  with MIMO radar: From co-existence to joint transmission},'' {\em IEEE
  Trans.\ Wireless Commun.}, vol.~17, no.~4, pp.~2755--2770, Apr. 2018.

\bibitem{wei2021orthogonal}
Z.~Wei, W.~Yuan, S.~Li, J.~Yuan, G.~Bharatula, R.~Hadani, and L.~Hanzo,
  ``Orthogonal time-frequency space modulation: A promising next-generation
  waveform,'' {\em IEEE Wireless Commun.}, vol.~28, no.~4, pp.~136--144, Aug.
  2021.

\bibitem{li2021performance}
S.~Li, J.~Yuan, W.~Yuan, Z.~Wei, B.~Bai, and D.~W.~K. Ng, ``Performance
  analysis of coded {OTFS} systems over high-mobility channels,'' {\em IEEE
  Trans.\ Wireless Commun.}, vol.~20, no.~9, pp.~6033--6048, Apr. 2021.

\bibitem{li2021hybrid}
S.~Li, W.~Yuan, Z.~Wei, J.~Yuan, B.~Bai, D.~W.~K. Ng, and Y.~Xie, ``{Hybrid MAP
  and PIC detection for OTFS modulation},'' {\em IEEE Trans.\ Veh.\ Technol.},
  vol.~70, no.~7, pp.~7193--7198, Jul. 2021.

\bibitem{li2021cross}
S.~Li, W.~Yuan, Z.~Wei, and J.~Yuan, ``Cross domain iterative detection for
  orthogonal time frequency space modulation,'' {\em IEEE Trans. Wireless
  Commun.}, vol.~21, pp.~2227--2242, Sept. Apr. 2022.

\bibitem{gaudio2020effectiveness}
L.~Gaudio, M.~Kobayashi, G.~Caire, and G.~Colavolpe, ``On the effectiveness of
  {OTFS} for joint radar parameter estimation and communication,'' {\em IEEE
  Trans.\ Wireless Commun.}, vol.~19, no.~9, pp.~5951--5965, Jun. 2020.

\bibitem{yuan2021integrated}
W.~Yuan, Z.~Wei, S.~Li, J.~Yuan, and D.~W.~K. Ng, ``Integrated sensing and
  communication-assisted orthogonal time frequency space transmission for
  vehicular networks,'' {\em IEEE J. Sel. Top. Signal Process.}, vol.~15,
  no.~6, pp.~1515--1528, Nov. 2021.

\bibitem{li2022novel}
S.~Li, W.~Yuan, C.~Liu, Z.~Wei, J.~Yuan, B.~Bai, and D.~W.~K. Ng, ``A novel
  {ISAC} transmission framework based on spatially-spread orthogonal time
  frequency space modulation,'' {\em IEEE J. Sel. Areas Commun.}, vol.~40,
  pp.~1854--1872, Jun. 2022.

\bibitem{liu2020radar}
F.~Liu, W.~Yuan, C.~Masouros, and J.~Yuan, ``Radar-assisted predictive
  beamforming for vehicular links: Communication served by sensing,'' {\em IEEE
  Trans.\ Wireless Commun.}, vol.~19, no.~11, pp.~7704--7719, Nov. 2020.

\bibitem{yuan2020bayesian}
W.~Yuan, F.~Liu, C.~Masouros, J.~Yuan, D.~W.~K. Ng, and
  N.~Gonz{\'a}lez-Prelcic, ``Bayesian predictive beamforming for vehicular
  networks: A low-overhead joint radar-communication approach,'' {\em IEEE
  Trans.\ Wireless Commun.}, vol.~20, no.~3, pp.~1442--1456, Mar. 2021.

\bibitem{liu2020deepresidual}
C.~Liu, X.~Liu, D.~W.~K. Ng, and J.~Yuan, ``Deep residual learning for channel
  estimation in intelligent reflecting surface-assisted multi-user
  communications,'' {\em IEEE Trans. Wireless Commun.}, vol.~21, no.~2,
  pp.~898--912, Feb. 2022.

\bibitem{xie2020unsupervised}
J.~Xie, J.~Fang, C.~Liu, and L.~Yang, ``Unsupervised deep spectrum sensing: A
  variational auto-encoder based approach,'' {\em IEEE Trans. Veh. Technol.},
  vol.~69, no.~5, pp.~5307--5319, May 2020.

\bibitem{liu2020deeptransfer}
C.~Liu, Z.~Wei, D.~W.~K. Ng, J.~Yuan, and Y.-C. Liang, ``Deep transfer learning
  for signal detection in ambient backscatter communications,'' {\em IEEE
  Trans. Wireless Commun.}, vol.~20, no.~3, pp.~1624--1638, Mar. 2021.

\bibitem{marzetta2016fundamentals}
T.~L. Marzetta, {\em Fundamentals of massive {MIMO}}.
\newblock Cambridge University Press, 2016.

\bibitem{barneto2021full}
C.~B. Barneto, S.~D. Liyanaarachchi, M.~Heino, T.~Riihonen, and M.~Valkama,
  ``Full duplex radio/radar technology: The enabler for advanced joint
  communication and sensing,'' {\em IEEE Wireless Commun.}, vol.~28, no.~1,
  pp.~82--88, Feb. 2021.

\bibitem{liu2014maximum}
C.~Liu and M.~Jin, ``Maximum-minimum spatial spectrum detection for cognitive
  radio using parasitic antenna arrays,'' in {\em Proc. IEEE Int. Conf. Commun.
  China (ICCC)}, pp.~365--369, Shanghai, China, Oct. 2014.

\bibitem{Niu2015a}
Y.~Niu, Y.~Li, D.~Jin, L.~Su, and A.~V. Vasilakos, ``{A survey of millimeter
  wave communications (mmWave) for 5G: opportunities and challenges},'' {\em
  Wireless Netw.}, vol.~21, no.~8, pp.~2657--2676, Nov. 2015.

\bibitem{zhao2018max}
W.~Zhao, C.~Liu, W.~Liu, and M.~Jin, ``Maximum eigenvalue based radar signal
  detection method for k distribution sea clutter environment,'' {\em Journal
  of Electronics $\&$ Information Technology}, vol.~40, no.~9, pp.~2235--2241,
  2018.

\bibitem{liu2016blind}
C.~Liu, H.~Li, and M.~Jin, ``Blind central-symmetry-based feature detection for
  spatial spectrum sensing,'' {\em IEEE Trans.\ Veh.\ Technol.}, vol.~65,
  no.~12, pp.~10147--10152, Dec.~2016.

\bibitem{zhao2018maximum}
W.~Zhao, C.~Liu, W.~Liu, and M.~Jin, ``Maximum eigenvalue-based target
  detection for the {K}-distributed clutter environment,'' {\em IET Radar,
  Sonar $\&$ Navigation}, vol.~12, no.~11, pp.~1294--1306, 2018.

\bibitem{liu2014blind}
C.~Liu, M.~Li, and M.-L. Jin, ``Blind energy-based detection for spatial
  spectrum sensing,'' {\em IEEE Wireless Commun. Lett.}, vol.~4, no.~1,
  pp.~98--101, Feb.~2015.

\bibitem{Trees2004optimum}
H.~L. Van~Trees, {\em Optimum array processing: {Part IV} of detection,
  estimation, and modulation theory}.
\newblock John Wiley $\&$ Sons, 2004.

\bibitem{Gross2015smart}
F.~B. Gross, {\em Smart Antennas with MATLAB}.
\newblock McGraw-Hill Education, 2015.

\bibitem{liu2022learning}
C.~Liu, W.~Yuan, S.~Li, X.~Liu, H.~Li, D.~W.~K. Ng, and Y.~Li, ``Learning-based
  predictive beamforming for integrated sensing and communication in vehicular
  networks,'' {\em IEEE J. Sel. Areas Commun.}, vol.~40, no.~8, pp.~2317--2334,
  Aug. 2022.

\bibitem{wymeersch20175g}
H.~Wymeersch, G.~Seco-Granados, G.~Destino, D.~Dardari, and F.~Tufvesson,
  ``{5G} mmwave positioning for vehicular networks,'' {\em IEEE Wireless
  Commun.}, vol.~24, no.~6, pp.~80--86, Dec. 2017.

\bibitem{zeng2019axxessing}
Y.~Zeng, Q.~Wu, and R.~Zhang, ``Accessing from the sky: A tutorial on uav
  communications for 5g and beyond,'' {\em Proc. IEEE}, vol.~107, no.~12,
  pp.~2327--2375, Dec. 2019.

\bibitem{liu2019deep}
C.~Liu, J.~Wang, X.~Liu, and Y.-C. Liang, ``Deep {CM-CNN} for spectrum sensing
  in cognitive radio,'' {\em IEEE J. Sel. Areas Commun.}, vol.~37, no.~10,
  pp.~2306--2321, Oct.~2019.

\bibitem{Goodfellow2016deep}
I.~Goodfellow, Y.~Bengio, A.~Courville, and Y.~Bengio, {\em Deep learning}.
\newblock MIT Press Cambridge, 2016.

\bibitem{liu2021location}
C.~Liu, W.~Yuan, Z.~Wei, X.~Liu, and D.~W.~K. Ng, ``Location-aware predictive
  beamforming for {UAV} communications: A deep learning approach,'' {\em IEEE
  Wireless Commun. Lett.}, vol.~10, no.~3, pp.~668--672, Mar. 2021.

\bibitem{lxm2021deep}
X.~Liu, C.~Liu, Y.~Li, B.~Vucetic, and D.~W.~K. Ng, ``Deep residual
  learning-assisted channel estimation in ambient backscatter communications,''
  {\em IEEE Wireless Commun. Lett.}, vol.~10, no.~2, pp.~339--343, Feb.~2021.

\bibitem{xie2019activity}
J.~Xie, C.~Liu, Y.-C. Liang, and J.~Fang, ``Activity pattern aware spectrum
  sensing: A {CNN}-based deep learning approach,'' {\em IEEE Commun. Lett.},
  vol.~23, no.~6, pp.~1025--1028, Jun.~2019.

\bibitem{liu2019maximum}
C.~Liu, J.~Wang, X.~Liu, and Y.-C. Liang, ``Maximum eigenvalue-based
  goodness-of-fit detection for spectrum sensing in cognitive radio,'' {\em
  IEEE Trans.\ Veh.\ Technol.}, vol.~68, no.~8, pp.~7747--7760, Aug.~2019.

\bibitem{liu2017optimal}
C.~Liu, H.~Li, J.~Wang, and M.~Jin, ``Optimal eigenvalue weighting detection
  for multi-antenna cognitive radio networks,'' {\em IEEE Trans.\ Wireless
  Commun.}, vol.~16, no.~4, pp.~2083--2096, Apr.~2017.

\bibitem{hu2021robust}
S.~Hu, Z.~Wei, Y.~Cai, C.~Liu, D.~W.~K. Ng, and J.~Yuan, ``Robust and secure
  sum-rate maximization for multiuser miso downlink systems with
  self-sustainable {IRS},'' {\em IEEE Trans.\ Commun.}, vol.~69, no.~10,
  pp.~7032--7049, Oct. 2021.

\bibitem{li2022how}
D.~Li, ``How many reflecting elements are needed for energy-and
  spectral-efficient intelligent reflecting surface-assisted communication,''
  {\em IEEE Trans.\ Commun.}, vol.~70, no.~2, pp.~1320--1331, Feb. 2022.

\bibitem{ye2021delay}
Y.~Ye, L.~Shi, X.~Chu, D.~Li, and G.~Lu, ``Delay minimization in wireless
  powered mobile edge computing with hybrid backcom and at,'' {\em IEEE
  Wireless Commun. Lett.}, vol.~10, no.~7, pp.~1532--1536, Jul. 2021.

\end{thebibliography}

\end{document}